\documentclass[10pt,conference,review]{IEEEtran}
\IEEEoverridecommandlockouts
  
\usepackage{amsmath,amssymb,amsfonts}
\usepackage{graphicx}
\usepackage{textcomp}
\usepackage{xcolor}
\usepackage{cite}
\usepackage{adjustbox}
\usepackage{color, colortbl}
\usepackage{calc}
\usepackage{balance}

\usepackage{subfigure}

\usepackage{amsmath}
\usepackage{pgfplots}

\pgfplotsset{compat=1.10}
\usepgfplotslibrary{fillbetween}
\def\BibTeX{{\rm B\kern-.05em{\sc i\kern-.025em b}\kern-.08emT\kern-.1667em\lower.7ex\hbox{E}\kern-.125emX}}

\usepackage{array}    
\usepackage{booktabs} 
\usepackage[skins]{tcolorbox}
\usepackage{tabularx}
\usepackage{multirow}
\usepackage[flushleft]{threeparttable}
\usepackage{amssymb}
\usepackage{enumitem}
\usepackage{rotate}
\usepackage{pgf}
\usepackage{graphicx}
\usepackage{colortbl}
\usepackage{arydshln}
\usepackage{dblfloatfix} 
\usepackage{times}
\usepackage{rotating}
\usepackage{makecell} 
\usepackage{tabularx}
\usepackage{balance}
\usepackage{booktabs}
\usepackage{wrapfig}
\usepackage{tikz}
\usetikzlibrary{angles}
\usepackage{makecell}
\usepackage{tabu}
\newcommand{\bi}{\begin{itemize}}
\newcommand{\ei}{\end{itemize}}
\newcommand{\be}{\begin{enumerate}}
\newcommand{\ee}{\end{enumerate}}

\usepackage{amsmath}
\usepackage{subfigure}
\usepackage{tablefootnote}

\usepackage{url}

\usepackage{tikz}
\usetikzlibrary{arrows.meta}
\tikzset{%
  >={Latex[width=2mm,length=2mm]},
            base/.style = {rectangle, rounded corners, draw=black,
                           minimum width=2.5cm, minimum height=1cm,
                           text centered, font=\sffamily},
  activityStarts/.style = {base, fill=blue!30},
       startstop/.style = {base, fill=red!30},
    activityRuns/.style = {base, fill=green!30},
         process/.style = {base, minimum width=2.5cm, fill=orange!15,
                           font=\ttfamily},
}

\makeatletter
\def\BState{\State\hskip-\ALG@thistlm}
\makeatother

\newcommand\MyBox[2]{
  \fbox{\lower0.75cm
    \vbox to 1.7cm{\vfil
      \hbox to 1.7cm{\hfil\parbox{1.4cm}{#1\\#2}\hfil}
      \vfil}%
  }%
}

\usepackage[final]{listings}
\usepackage{algorithm}
\usepackage[noend]{algpseudocode}

\usepackage[framemethod=tikz]{mdframed}
\usetikzlibrary{shadows}
\usepackage{graphics}
\newmdenv[
tikzsetting= {fill=gray!10},
linewidth=1pt,
roundcorner=2pt, 
shadow=false
]{myshadowbox}

\begin{document}

\title{Software Engineering for Fairness: \\A Case Study with Hyperparameter Optimization}

\author{\IEEEauthorblockN{Blinded for review}}

\author{\IEEEauthorblockN{Joymallya Chakraborty, Tianpei Xia, Fahmid M. Fahid, Tim Menzies}
\IEEEauthorblockA{jchakra@ncsu.edu,txia4@ncsu.edu,ffahid@ncsu.edu,timm@ieee.org\\
North Carolina State University\\}}

\maketitle
\thispagestyle{plain}
\pagestyle{plain}

\begin{abstract}
We assert that it is the ethical duty of software engineers to strive to reduce software discrimination.  This paper discusses how that might be done.

This is an important topic since
machine learning software is increasingly being used to make decisions that affect people's lives.  Potentially,  the application of that software will result in fairer decisions because (unlike humans) machine learning software is not biased. However, recent results show that the software within many data mining packages exhibit ``group discrimination''; i.e. their decisions are  inappropriately affected by   “protected attributes” (e.g., race, gender, age, etc.).

There has been much prior work  on validating the fairness of machine-learning models (by recognizing when such software discrimination exists). But after detection, comes mitigation. What steps can ethical software engineers take to reduce discrimination in the software they produce?  

This paper shows that making \textit{fairness} as a goal during hyperparamter optimization can (a) preserve the predictive power of a model learned from a data miner while also (b) generates fairer results. To the best of  our knowledge, this is the first application  of hyperparameter optimization as a tool for software engineers to generate fairer software.

\end{abstract}


\section{Introduction}
Many high-stake applications such as finance, hiring, admissions, criminal justice use algorithmic decision-making frequently. In some cases, machine learning models make better decisions than human can do \cite{Brun:2018:SF:3236024.3264838,Aydemir:2018:RES:3194770.3194778}. But there are many scenarios where machine learning software has been found to be biased and generating arguably unfair decisions. Google's sentiment analyzer model which determines positive or negative sentiment, gives negative score to the sentences such as \textit{`I am a Jew', and `I am homosexual'}\cite{Google_Sentiment}. Facial recognition software which predicts characteristics such as gender, age from images has been found to have a much higher error rate for dark-skinned women compared to light-skinned men \cite{Gender_Bias}. A popular photo tagging model has assigned animal category labels to dark skinned people \cite{Google_Photo}. Recidivism assessment models used by the criminal justice system have been found to be more likely to falsely label black defendants as future criminals at almost twice the rate as white defendants \cite{Machine_Bias}. Amazon.com stopped using automated job recruiting model after detection of bias against women\cite{Amazon_Bias}. Cathy O'Neil’ provided even more examples of unfair decisions made by software in her book ``Weapons of Math Destruction''\cite{O'Neil:2016:WMD:3002861}. She argued that machine learning software generates models that are full of bias. Hence, this is one of the reasons their application results in unfair decisions.

Machine learning software, by its nature, is always a form of statistical discrimination. The discrimination becomes objectionable when it places certain privileged groups at systematic advantage and certain unprivileged groups at systematic disadvantage. In certain situations, such as employment (hiring and firing), discrimination is not only objectionable, but illegal.

Issues of \textit{fairness} have been explored in many  recent papers in the SE research literature. Angell et al. \cite{Angell:2018:TAT:3236024.3264590}  commented that issues of fairness are analogous to other measures of software quality. Galhotra and his colleagues discussed how to efficiently generate test cases to test for discrimination\cite{Galhotra_2017}. Udeshi et al. \cite{Udeshi_2018} worked on generating discriminatory  inputs for machine learning software. Albarghouthi et al. \cite{Albarghouthi:2019:FP:3287560.3287588} explored if fairness can be wired into annotations within a program while Tramer et al. proposed different ways to measure discrimination \cite{Tramer_2017}.

All the above SE research detects unfairness. Our work takes a step further and asks how to mitigate
unfairness. We propose that every machine learning model must go through fairness testing phase before it is applied. If bias is found, then the model needs to be optimized. Hence, we have converted ``discrimination problem'' into an optimization problem. We think that if \textit{fairness} becomes a goal while learning, then the  models created in that way will generate fairer results. In this study, we investigated whether model parameter tuning can help us to make the model fair or not. 
 
 In machine learning, many  {\em hyperparameters} control   inductive process ; e.g. the   `splitter' of CART~\cite{breiman2017classification}.  They are very important because they directly control the behaviors of the training algorithm and impact the performance of the model. Therefore, the selection of appropriate parameters plays a critical role in the performance of machine learning models. Our study applies \textit{hyperparameter optimization} to make a model fair without losing predictive power. So, it becomes \textit{multiobjective optimization} problem as we are dealing with more than one objective. 
 

\section{But is this a Problem for Software Engineers?} 
We are not the only ones to assert that software fairness is a concern that must be addressed 
by software engineers. Other SE researchers are also exploring this issues \cite{Angell:2018:TAT:3236024.3264590,Brun:2018:SF:3236024.3264838,Aydemir:2018:RES:3194770.3194778}. For example,   IEEE/ACM 
recently organized a workshop on software fairness called \textit{Fairware 2018}\footnote{http://fairware.cs.umass.edu/}.

Nevertheless,   when discussing this work with colleagues, we are still sometimes
asked if this problem {\em should}
or {\em can} be solved by 
software engineers. We reply that:
\bi
\item
It {\em should} be the goal of  software developers to ensure
that software conforms to  its
required ethical standards.
\item
Even if we think that  fairness is not our
problem, our users may disagree.
When users discover problems with software, it is the job of
the person maintaining that software (i.e.  a software engineer) to fix that problem.
\item
Further, we also think that this problem {\em can} be solved by software engineers. Hyperparameter optimization is now a standard tool
in software analytics~\cite{xia2018hyperparameter,osman2017hyperparameter}. What we are arguing here is that now that those same tools, that  have been matured within the SE community (by SE researchers and practitioners),
can now be applied to other problems (e.g.  as discussed in this paper, how to  mitigate unfair software).
\ei

\section{Terminology}

We say that a label is called \textit{favorable label} if its  value corresponds to an outcome that gives an advantage to the receiver. Examples like - being hired for a job, receiving a loan. \textit{Protected attribute} is an attribute that divides a population into two groups that have difference in terms of benefit received. Like - sex, race. These attributes are not universal, but are specific to application. \textit{Group fairness} is the goal that based on the protected attribute, privileged and unprivileged groups will be treated similarly. \textit{Individual fairness} is the goal of similar individuals will receive similar outcomes.  Our paper studies Group fairness only.
By definition, ``Bias is a systematic error '' \cite{bias_systemetic}. Our main concern is unwanted bias that puts privileged groups at a systematic advantage and unprivileged groups at a systematic disadvantage. A \textit{fairness metric} is a quantification of unwanted bias in models or training data \cite{IBM}. We used two such fairness metrics in our experiment-

\bi
\item \textbf{Equal Opportunity Difference(EOD)}:  Delta in true positive rates in unprivileged and privileged groups \cite{IBM}. 
\item \textbf{Average Odds Difference(AOD)}: Average delta in false positive rates and true positive rates between privileged and unprivileged groups \cite{IBM}.
\ei
Both are computed using the input and output datasets to a classifier. A value of 0 implies that both groups have equal benefit, a value lesser than 0 implies higher benefit for the privileged group and a value greater than 0 implies higher benefit for the unprivileged group. In this study, we have taken absolute value of these metrics. 

\section{Methodology}

\subsection{Hyperparameter Optimization}
 Hyperparameter optimization is the process of searching the most optimal hyperparameters in machine learning learners~\cite{biedenkapp2018hyperparameter}~\cite{franceschi2017forward}. There are four common algorithms: grid search, random search, Bayesian optimization and SMBO.

\textit{Grid search}~\cite{bergstra2011algorithms} implements all possible combination of hyperparameters for a learner and tries to find out the best one. It suffers if data have high dimensional space called the ``curse of dimensionality''. It tries all combinations but only a few of the tuning parameters really matter~\cite{bergstra2012random}.

\textit{Random search}~\cite{bergstra2012random} sets up a grid of hyperparameter values and select random combinations to train the model and evaluate. The evaluation is based on a specified probability distribution. The main problem of this method is at each step, it does not use information from the prior steps. 

In contrast to Grid or Random search, \textit{Bayesian optimization}~\cite{pelikan1999boa} keeps track of past evaluation results and use them to build a probabilistic model mapping hyperparameters to a probability of a score on the objective function \cite{Will_Koehrsen}. This probabilistic model is called ``surrogate'' for the objective function. The idea is to find the next set of hyperparameters to evaluate on the actual objective function by selecting hyperparameters that perform best on the surrogate function.

\textit{Sequential model-based optimization (SMBO)} \cite{10.1007/978-3-642-25566-3_40} is a formalization of Bayesian optimization. It runs trials one by one sequentially, each time trying better hyperparameters using Bayesian reasoning and updating the surrogate model \cite{Will_Koehrsen}.

Recent studies have shown that hyperparameter optimization can achieve better performance than using ``off-the-shelf'' configurations in several research areas in software engineering, e.g., software effort estimation\cite{xia2018hyperparameter} and software defect prediction\cite{osman2017hyperparameter}. We are first to apply hyperparameter optimization in software fairness domain.

\subsection{FLASH: A Fast Sequential Model-Based Method}
Nair et al. \cite{8469102} proposed a fast SMBO approach called FLASH for multiobjective optimization. FLASH's acquisition function uses Maximum Mean. Maximum Mean returns the sample (configuration) with the highest expected (performance) measure. FLASH models each objective as a separate performance (CART) model. Because the CART model can be trained for one performance measure or dependent value. Nair reports that FLASH runs orders of magnitude faster than NSGA-II, but that was for software configuration problems. This work is the first study to try using  FLASH to optimize for learner performance while at the same time improving fairness.

\begin{table}[]
\scriptsize
\caption{The Description of Datasets used in our study, N=\#rows. F=\#features, FAV=favorable.
``recid''=recidivate}
\label{tbl:dataset}
\begin{tabular}{|l@{~}|l@{~}|l@{~}|l@{~}|l@{~}|p{0.7cm}@{~}|p{0.6cm}|}
\hline
\rowcolor[HTML]{C0C0C0} 
\multicolumn{1}{|c|}{\cellcolor[HTML]{C0C0C0}} & \multicolumn{1}{c|}{\cellcolor[HTML]{C0C0C0}} & \multicolumn{1}{c|}{\cellcolor[HTML]{C0C0C0}} & \multicolumn{2}{c|}{\cellcolor[HTML]{C0C0C0}Protected Attribute} & \multicolumn{2}{c|}{\cellcolor[HTML]{C0C0C0}Label} \\ \cline{4-7} 
\rowcolor[HTML]{C0C0C0} 
\multicolumn{1}{|c|}{\multirow{-2}{*}{\cellcolor[HTML]{C0C0C0}Dataset}} & \multicolumn{1}{c|}{\multirow{-2}{*}{\cellcolor[HTML]{C0C0C0}N}} & \multicolumn{1}{c|}{\multirow{-2}{*}{\cellcolor[HTML]{C0C0C0}F}} & \multicolumn{1}{c|}{\cellcolor[HTML]{C0C0C0}Privileged} & \multicolumn{1}{c|}{\cellcolor[HTML]{C0C0C0}Unprivileged} & \multicolumn{1}{c|}{\cellcolor[HTML]{C0C0C0}Fav} & \multicolumn{1}{c|}{\cellcolor[HTML]{C0C0C0}UnFav} \\ \hline
\begin{tabular}[c]{@{}l@{}}Adult \\ Census\\ Income\tablefootnote{https://archive.ics.uci.edu/ml/datasets/adult}\end{tabular} & 48,842 & 14 & \begin{tabular}[c]{@{}l@{}}Sex - Male\\ Race - White\end{tabular} & \begin{tabular}[c]{@{}l@{}}Sex - Female\\ Race - Non-\\ white\end{tabular} & \begin{tabular}[c]{@{}l@{}}High \\ Income\end{tabular} & \begin{tabular}[c]{@{}l@{}}Low \\ Income\end{tabular} \\ \hline
Compas\tablefootnote{https://github.com/propublica/compas-analysis} & 7,214 & 28 & \begin{tabular}[c]{@{}l@{}}Sex - Female\\ Race - Caucasian\end{tabular} & \begin{tabular}[c]{@{}l@{}}Sex - Male\\ Race - Not \\ Caucasian\end{tabular} & \begin{tabular}[c]{@{}l@{}}Did \\ recid\end{tabular} & \begin{tabular}[c]{@{}l@{}}Did \\ not \\ recid\end{tabular} \\ \hline
\begin{tabular}[c]{@{}l@{}}German\\ Credit \\ Data\tablefootnote{https://archive.ics.uci.edu/ml/datasets/statlog+(german+credit+data}\end{tabular} & 1,000 & 20 & \begin{tabular}[c]{@{}l@{}}Sex - Male\\ Age - Old\end{tabular} & \begin{tabular}[c]{@{}l@{}}Sex - Female\\ Age - Young\end{tabular} & Good Credit & Bad Credit \\ \hline
\end{tabular}
\footnotetext{https://archive.ics.uci.edu/ml/datasets/statlog+(german+credit+data)}
\end{table}

\section{Results}

\newenvironment{RQ}{\vspace{2mm}\begin{tcolorbox}[enhanced,width=3.4in,size=fbox,fontupper=\small,colback=blue!5,drop shadow southwest,sharp corners]}{\end{tcolorbox}}

\subsection{
RQ1: Does optimizing for fairness damage model prediction performance ?}

We have verified our method along with four other related works to answer this question. Table \ref{tbl:dataset} shows the datasets we used. We randomly divided them into three sets - training (70\%), validation (15\%) and test (15\%). Prior researchers who worked with these datasets have used \textit{Logistic Regression} as classification model \cite{Kamishima,NIPS2017_6988,Hardt}. We also decided to use this learner. Before moving to results, here we briefly describe prior works which we selected for our study. There are mainly three kinds of prior works -

\bi
\item \textbf{Pre-processing algorithms}: In this method, data is pre-processed(before classification) in such a way that discrimination is reduced. Kamiran et al. proposed \textit{Reweighing} \cite{Kamiran2012} method that generates weights for the training examples in each (group, label) combination differently to ensure fairness. Later, Calmon et al. proposed an \textit{Optimized pre-processing} method \cite{NIPS2017_6988} which learns a probabilistic transformation that edits the labels and features with individual distortion and group fairness.

\item \textbf{In-processing algorithms}: This is an optimization approach where dataset is divided into train, validation and test set. After learning from training data, model is optimized on the validation set and finally applied on the test set. Our \textit{Hyperparameter Optimization} using FLASH approach lies into this category. Zhang et al. proposed \textit{Adversarial debiasing}  \cite{Zhang:2018:MUB:3278721.3278779} method which learns a classifier to maximize accuracy and simultaneously reduce an adversary's ability to determine the protected attribute from the predictions. This generates a fair classifier because the predictions cannot carry any group discrimination information that the adversary can exploit.

\item \textbf{Post-processing algorithms}: Hereafter classification, the class labels are changed to reduce discrimination. Kamiran et al. proposed \textit{Reject option classification} approach \cite{Kamiran:2018:ERO:3165328.3165686} which gives unfavorable outcomes to privileged groups and favorable outcomes to unprivileged groups within a confidence band around the decision boundary with the highest uncertainty.

\ei

\begin{table}[]
\centering
\scriptsize
\caption{Optimizing just for fairness.  
Change in  Recall and False alarm before and after bias mitigation. Gray= improvement; black= damage.}
\label{tbl:fairness_cost}
\begin{tabular}{|l|c|c|r|r|r|r|}
\hline
\rowcolor[HTML]{C0C0C0} 
\cellcolor[HTML]{C0C0C0} & \cellcolor[HTML]{C0C0C0} & \cellcolor[HTML]{C0C0C0} & \multicolumn{2}{c|}{\cellcolor[HTML]{C0C0C0}Recall} & \multicolumn{2}{c|}{\cellcolor[HTML]{C0C0C0}False alarm} \\ \cline{4-7} 
\rowcolor[HTML]{C0C0C0} 
\multirow{-2}{*}{\cellcolor[HTML]{C0C0C0}Algorithm} & \multirow{-2}{*}{\cellcolor[HTML]{C0C0C0}Dataset} & \multirow{-2}{*}{\cellcolor[HTML]{C0C0C0}\begin{tabular}[c]{@{}c@{}}Protected\\ Attribute\end{tabular}} & Before & After & Before & After \\ \hline
 &  & Sex & 0.38 & \cellcolor[HTML]{C0C0C0}0.43 & 0.06 & \cellcolor[HTML]{333333}{\color[HTML]{FFFFFF} 0.10} \\ \cline{3-7} 
 & \multirow{-2}{*}{Adult} & Race & 0.38 & \cellcolor[HTML]{C0C0C0}0.39 & 0.06 & 0.06 \\ \cline{2-7} 
 &  & Sex & 0.52 & \cellcolor[HTML]{C0C0C0}0.53 & 0.19 & \cellcolor[HTML]{333333}{\color[HTML]{FFFFFF} 0.22} \\ \cline{3-7} 
 & \multirow{-2}{*}{Compas} & Race & 0.52 & \cellcolor[HTML]{C0C0C0}0.57 & 0.19 & \cellcolor[HTML]{333333}{\color[HTML]{FFFFFF} 0.30} \\ \cline{2-7} 
 &  & Sex & 0.04 & 0.04 & 0.03 & \cellcolor[HTML]{333333}{\color[HTML]{FFFFFF} 0.06} \\ \cline{3-7} 
\multirow{-6}{*}{Reweighing} & \multirow{-2}{*}{German} & Age & 0.04 & \cellcolor[HTML]{C0C0C0}0.06 & 0.03 & \cellcolor[HTML]{C0C0C0}{\color[HTML]{333333} 0.00} \\ \hline
 &  & Sex & 0.40 & \cellcolor[HTML]{333333}{\color[HTML]{FFFFFF} 0.32} & 0.06 & \cellcolor[HTML]{333333}{\color[HTML]{FFFFFF} 0.07} \\ \cline{3-7} 
 & \multirow{-2}{*}{Adult} & Race & 0.40 & \cellcolor[HTML]{333333}{\color[HTML]{FFFFFF} 0.38} & 0.06 & \cellcolor[HTML]{333333}{\color[HTML]{FFFFFF} 0.07} \\ \cline{2-7} 
 &  & Sex & 0.52 & 0.52 & 0.19 & \cellcolor[HTML]{333333}{\color[HTML]{FFFFFF} 0.22} \\ \cline{3-7} 
 & \multirow{-2}{*}{Compas} & Race & 0.52 & 0.52 & 0.19 & \cellcolor[HTML]{333333}{\color[HTML]{FFFFFF} 0.22} \\ \cline{2-7} 
 &  & Sex & 0.04 & \cellcolor[HTML]{C0C0C0}0.36 & 0.03 & \cellcolor[HTML]{333333}{\color[HTML]{FFFFFF} 0.12} \\ \cline{3-7} 
\multirow{-6}{*}{\begin{tabular}[c]{@{}c@{}}Optimized\\ Pre-\\ processing\end{tabular}} & \multirow{-2}{*}{German} & Age & 0.04 & \cellcolor[HTML]{C0C0C0}0.10 & 0.03 & \cellcolor[HTML]{333333}{\color[HTML]{FFFFFF} 0.06} \\ \hline
 &  & Sex & 0.38 & \cellcolor[HTML]{C0C0C0}0.43 & 0.06 & \cellcolor[HTML]{333333}{\color[HTML]{FFFFFF} 0.09} \\ \cline{3-7} 
 & \multirow{-2}{*}{Adult} & Race & 0.37 & \cellcolor[HTML]{C0C0C0}0.38 & 0.06 & \cellcolor[HTML]{333333}{\color[HTML]{FFFFFF} 0.09} \\ \cline{2-7} 
 &  & Sex & 0.48 & \cellcolor[HTML]{C0C0C0}0.50 & 0.19 & \cellcolor[HTML]{333333}{\color[HTML]{FFFFFF} 0.20} \\ \cline{3-7} 
 & \multirow{-2}{*}{Compas} & Race & 0.48 & \cellcolor[HTML]{C0C0C0}0.58 & 0.18 & \cellcolor[HTML]{333333}{\color[HTML]{FFFFFF} 0.31} \\ \cline{2-7} 
 &  & Sex & 0.04 & \cellcolor[HTML]{C0C0C0}0.64 & 0.03 & \cellcolor[HTML]{333333}{\color[HTML]{FFFFFF} 0.71} \\ \cline{3-7} 
\multirow{-6}{*}{\begin{tabular}[c]{@{}c@{}}Adversial \\ Debiasing\end{tabular}} & \multirow{-2}{*}{German} & Age & 0.08 & \cellcolor[HTML]{C0C0C0}0.88 & 0.04 & \cellcolor[HTML]{333333}{\color[HTML]{FFFFFF} 0.85} \\ \hline
 &  & Sex & 0.38 & \cellcolor[HTML]{333333}{\color[HTML]{FFFFFF} 0.24} & 0.06 & \cellcolor[HTML]{C0C0C0}0.05 \\ \cline{3-7} 
 & \multirow{-2}{*}{Adult} & Race & 0.38 & \cellcolor[HTML]{333333}{\color[HTML]{FFFFFF} 0.04} & 0.06 & \cellcolor[HTML]{C0C0C0}0.04 \\ \cline{2-7} 
 &  & Sex & 0.52 & \cellcolor[HTML]{C0C0C0}0.97 & 0.19 & \cellcolor[HTML]{333333}{\color[HTML]{FFFFFF} 0.89} \\ \cline{3-7} 
 & \multirow{-2}{*}{Compas} & Race & 0.52 & \cellcolor[HTML]{C0C0C0}0.68 & 0.19 & \cellcolor[HTML]{333333}{\color[HTML]{FFFFFF} 0.38} \\ \cline{2-7} 
 &  & Sex & 0.04 & 0.04 & 0.03 & 0.03 \\ \cline{3-7} 
\multirow{-6}{*}{\begin{tabular}[c]{@{}c@{}}Reject\\ Option\end{tabular}} & \multirow{-2}{*}{German} & Age & 0.04 & 0.04 & 0.03 & 0.03 \\ \hline
\multicolumn{1}{|l|}{} &  & Sex & 0.42 & \cellcolor[HTML]{333333}{\color[HTML]{FFFFFF} 0.04} & 0.08 & \cellcolor[HTML]{C0C0C0}0.01 \\ \cline{3-7} 
\multicolumn{1}{|l|}{} & \multirow{-2}{*}{Adult} & Race & 0.42 & \cellcolor[HTML]{333333}{\color[HTML]{FFFFFF} 0.12} & 0.08 & \cellcolor[HTML]{C0C0C0}0.01 \\ \cline{2-7} 
\multicolumn{1}{|l|}{} &  & Sex & 0.52 & \cellcolor[HTML]{333333}{\color[HTML]{FFFFFF} 0.50} & 0.28 & \cellcolor[HTML]{333333}{\color[HTML]{FFFFFF} 0.31} \\ \cline{3-7} 
\multicolumn{1}{|l|}{} & \multirow{-2}{*}{Compas} & Race & 0.52 & \cellcolor[HTML]{333333}{\color[HTML]{FFFFFF} 0.50} & 0.28 & \cellcolor[HTML]{333333}{\color[HTML]{FFFFFF} 0.31} \\ \cline{2-7} 
\multicolumn{1}{|l|}{} &  & Sex & 0.13 & 0.13 & 0.03 & 0.03 \\ \cline{3-7} 
\multicolumn{1}{|l|}{\multirow{-6}{*}{\begin{tabular}[c]{@{}l@{}}FLASH \\ optimizes for\\ AOD \& EOD\end{tabular}}} & \multirow{-2}{*}{German} & Age & 0.13 & 0.13 & 0.03 & 0.03 \\ \hline
\end{tabular}
\end{table}

Table \ref{tbl:fairness_cost} shows the results of our approach (FLASH) and four algorithms from prior works. We see that there are a few gray cells and many black cells indicating that achieving fairness damages performance - which bolsters the conclusion made by Berk et al.\cite{berk2017convex}. In summary,  fairness can have a cost. Our next question checks if multiobjective optimization can better trade-off between performance and fairness.  

\subsection{RQ2: Can we optimize machine learning model for both fairness and performance?}

Here, we applied  FLASH algorithm but this time, we considered four goals together: \textit{recall, false alarm, AOD, EOD}. The first two are related to performance and second two are related to fairness. For
recall, {\em larger} values are {\em better} while for everything
else, {\em smaller} is {\em better}.
 For this part of our study, we used two learning models - logistic regression and CART. 
 
 We have chosen four hyperparameters for both the learners to optimize for. For logistic regression (C, penalty, solver, max\_iter) and for CART - (criterion, splitter , min\_samples\_leaf, min\_samples\_split). Table \ref{tbl:multiobjective_results} shows the results. The ``Before'' column shows results with no tuning and ``After'' column shows tuned results. We can see that for
 the German dataset, we improved three objectives and recall did not decrease. In the Adult dataset, we improved three objectives with minor damage of recall. 
 With the
 Compas dataset, there was no improvement. 
 
 In summary, the results are clearly indicating if  multiobjective  optimization understand {\em all} the goals of learning
 (fairness {\em and performance}), then it is possible to achieve one without
 damaging the other. Our last research question asks  what is the cost of this kind of optimization.

\begin{table*}[]
\centering
\footnotesize
\caption{Optimizing for fairness, lower false alarm and higher recall. Gray=improvement; black=damage.
Note that, compared to Table~\ref{tbl:fairness_cost}, there is far less damage.}
\label{tbl:multiobjective_results}
\begin{tabular}{|l|l|c|r|l|r|l|r|l|
>{\columncolor[HTML]{FFFFFF}}r |l|}
\hline
\rowcolor[HTML]{C0C0C0} 
\cellcolor[HTML]{C0C0C0} & \cellcolor[HTML]{C0C0C0} & \cellcolor[HTML]{C0C0C0} & \multicolumn{2}{c|}{\cellcolor[HTML]{C0C0C0}Recall} & \multicolumn{2}{c|}{\cellcolor[HTML]{C0C0C0}\begin{tabular}[c]{@{}c@{}}False\\ alarm\end{tabular}} & \multicolumn{2}{c|}{\cellcolor[HTML]{C0C0C0}AOD} & \multicolumn{2}{c|}{\cellcolor[HTML]{C0C0C0}EOD} \\ \cline{4-11} 
\rowcolor[HTML]{C0C0C0} 
\multirow{-2}{*}{\cellcolor[HTML]{C0C0C0}Model} & \multirow{-2}{*}{\cellcolor[HTML]{C0C0C0}Dataset} & \multirow{-2}{*}{\cellcolor[HTML]{C0C0C0}\begin{tabular}[c]{@{}c@{}}Protected\\ Attribute\end{tabular}} & Before & After & Before & After & Before & After & Before & After \\ \hline
 &  & Sex & 0.42 & \cellcolor[HTML]{333333}{\color[HTML]{FFFFFF} 0.04} & 0.08 & \cellcolor[HTML]{C0C0C0}0.01 & 0.31 & \cellcolor[HTML]{C0C0C0}0.03 & 0.49 & \cellcolor[HTML]{C0C0C0}0.05 \\ \cline{3-11} 
 & \multirow{-2}{*}{Adult} & Race & 0.42 & \cellcolor[HTML]{333333}{\color[HTML]{FFFFFF} 0.12} & 0.08 & \cellcolor[HTML]{C0C0C0}0.01 & 0.14 & \cellcolor[HTML]{C0C0C0}0.05 & 0.21 & \cellcolor[HTML]{C0C0C0}0.03 \\ \cline{2-11} 
 &  & Sex & 0.52 & 0.52 & 0.28 & \cellcolor[HTML]{333333}{\color[HTML]{FFFFFF} 0.31} & 0.22 & 0.22 & 0.27 & 0.27 \\ \cline{3-11} 
 & \multirow{-2}{*}{Compas} & Race & 0.52 & 0.52 & 0.28 & \cellcolor[HTML]{333333}{\color[HTML]{FFFFFF} 0.31} & 0.25 & \cellcolor[HTML]{C0C0C0}0.15 & 0.34 & \cellcolor[HTML]{C0C0C0}0.23 \\ \cline{2-11} 
 &  & Sex & 0.13 & 0.13 & 0.03 & 0.03 & 0.16 & 0.16 & 0.03 & 0.03 \\ \cline{3-11} 
\multirow{-6}{*}{\begin{tabular}[c]{@{}c@{}}Logistic\\ regression\end{tabular}} & \multirow{-2}{*}{German} & Age & 0.13 & 0.13 & 0.03 & 0.03 & 0.14 & 0.14 & 0.05 & 0.05 \\ \hline
 &  & Sex & 0.41 & 0.41 & 0.07 & 0.07 & 0.30 & 0.30 & 0.49 & 0.49 \\ \cline{3-11} 
 & \multirow{-2}{*}{Adult} & Race & 0.41 & 0.41 & 0.07 & 0.07 & 0.14 & 0.14 & 0.22 & 0.22 \\ \cline{2-11} 
 &  & Sex & 0.53 & 0.53 & 0.25 & 0.25 & 0.19 & 0.19 & 0.20 & 0.20 \\ \cline{3-11} 
 & \multirow{-2}{*}{Compas} & Race & 0.53 & 0.53 & 0.25 & 0.25 & 0.10 & 0.10 & 0.13 & 0.13 \\ \cline{2-11} 
 &  & Sex & 0.13 & 0.13 & 0.05 & \cellcolor[HTML]{C0C0C0}0.03 & 0.20 & \cellcolor[HTML]{C0C0C0}0.18 & 0.10 & \cellcolor[HTML]{C0C0C0}0.07 \\ \cline{3-11} 
\multirow{-6}{*}{CART} & \multirow{-2}{*}{German} & Age & 0.13 & 0.13 & 0.05 & \cellcolor[HTML]{C0C0C0}0.03 & 0.13 & \cellcolor[HTML]{C0C0C0}0.11 & 0.03 & \cellcolor[HTML]{C0C0C0}0.03 \\ \hline
\end{tabular}
\end{table*}

\subsection{ RQ3. How much time does optimization take?}

Default logistic regression takes 0.56s, 0.15s and 0.11s for Adult, Compas and German dataset respectively. When we apply hyperparameter optimization, the cumulative time for training, tuning and testing become 16.33s, 4.34s and 3.55s for those datasets. 
We assert that  runtimes of less than 20 seconds is a relatively small price to pay to ensure fairness. 

As to larger, more complex problems, Nair et al.~\cite{8469102} reports
that FLASH scales to problems with larger order of magnitude than other optimizers. It is a matter for future research to see if such scale is possible/required to handle fairness of SE data.

\section{Conclusion \& Future Work} Our experiments show that it might be  possible to make software fair, without compromising other design goals (like predictive performance). Like Brun et al. \cite{Brun:2018:SF:3236024.3264838}, we   propose that software bias detection and mitigation should be included in the software life-cycle. In agile practices, before any release, software should go through fairness testing and mitigation.

In this study, we only considered logistic regression and CART decision tree. In the future, we will explore more learning models. Another area to explore in the future is more data sets.
Here, we used the same three datasets used by other  publications in this area. All these datasets are small and so may not be representative of 
other real world scenarios. Ideally,  software companies should consider making  their data available which they think might be beneficial for bias related study. Data is not the only challenge,  domain knowledge is important to understand the significance of protected attributes. For example,
can/should we  train our model without
any protected attributes at all (gender, race, age)? In our experience,  domain experts have strong opinions
on that matter.

\balance
\bibliographystyle{IEEEtran}
\bibliography{bibliography}

\begin{thebibliography}{10}
\providecommand{\url}[1]{#1}
\csname url@samestyle\endcsname
\providecommand{\newblock}{\relax}
\providecommand{\bibinfo}[2]{#2}
\providecommand{\BIBentrySTDinterwordspacing}{\spaceskip=0pt\relax}
\providecommand{\BIBentryALTinterwordstretchfactor}{4}
\providecommand{\BIBentryALTinterwordspacing}{\spaceskip=\fontdimen2\font plus
\BIBentryALTinterwordstretchfactor\fontdimen3\font minus
  \fontdimen4\font\relax}
\providecommand{\BIBforeignlanguage}[2]{{%
\expandafter\ifx\csname l@#1\endcsname\relax
\typeout{** WARNING: IEEEtran.bst: No hyphenation pattern has been}%
\typeout{** loaded for the language `#1'. Using the pattern for}%
\typeout{** the default language instead.}%
\else
\language=\csname l@#1\endcsname
\fi
#2}}
\providecommand{\BIBdecl}{\relax}
\BIBdecl

\bibitem{Brun:2018:SF:3236024.3264838}
Y.~Brun and A.~Meliou, ``Software fairness,'' ser. ESEC/FSE 18.\hskip 1em plus
  0.5em minus 0.4em\relax NY, USA: ACM, pp. 754--759.

\bibitem{Aydemir:2018:RES:3194770.3194778}
F.~B. Aydemir and F.~Dalpiaz, ``A roadmap for ethics-aware software
  engineering,'' ser. FairWare '18.\hskip 1em plus 0.5em minus 0.4em\relax NY,
  USA: ACM, pp. 15--21.

\bibitem{Google_Sentiment}
\BIBentryALTinterwordspacing
``Google’s sentiment analyzer thinks being gay is bad,'' \emph{Motherboard},
  Oct 2017. [Online]. Available: \url{https://bit.ly/2yMax8V}
\BIBentrySTDinterwordspacing

\bibitem{Gender_Bias}
\BIBentryALTinterwordspacing
``Study finds gender and skin-type bias in commercial artificial-intelligence
  systems.'' [Online]. Available: \url{https://bit.ly/2LxosK6}
\BIBentrySTDinterwordspacing

\bibitem{Google_Photo}
\BIBentryALTinterwordspacing
``Google apologizes for mis-tagging photos of african americans,'' July 2015.
  [Online]. Available: \url{https://cbsn.ws/2LBYbdy}
\BIBentrySTDinterwordspacing

\bibitem{Machine_Bias}
\BIBentryALTinterwordspacing
``Machine bias,'' \emph{www.propublica.org}, May 2016. [Online]. Available:
  \url{https://www.propublica.org/article/machine-bias-risk-assessments-in-criminal-sentencing}
\BIBentrySTDinterwordspacing

\bibitem{Amazon_Bias}
\BIBentryALTinterwordspacing
``Amazon scraps secret ai recruiting tool that showed bias against women,'' Oct
  2018. [Online]. Available: \url{https://reut.rs/2Po4ZJi}
\BIBentrySTDinterwordspacing

\bibitem{O'Neil:2016:WMD:3002861}
C.~O'Neil, \emph{Weapons of Math Destruction: How Big Data Increases Inequality
  and Threatens Democracy}.\hskip 1em plus 0.5em minus 0.4em\relax Crown
  Publishing Group, 2016.

\bibitem{Angell:2018:TAT:3236024.3264590}
R.~Angell, B.~Johnson, Y.~Brun, and A.~Meliou, ``Themis: Automatically testing
  software for discrimination,'' ser. ESEC/FSE 18.

\bibitem{Galhotra_2017}
S.~Galhotra, Y.~Brun, and A.~Meliou, ``Fairness testing: testing software for
  discrimination,'' \emph{ESEC/FSE 17}.

\bibitem{Udeshi_2018}
S.~Udeshi, P.~Arora, and S.~Chattopadhyay, ``Automated directed fairness
  testing,'' \emph{ASE18}.

\bibitem{Albarghouthi:2019:FP:3287560.3287588}
A.~Albarghouthi and S.~Vinitsky, ser. FAT* '19.\hskip 1em plus 0.5em minus
  0.4em\relax NY, USA: ACM, pp. 211--219.

\bibitem{Tramer_2017}
F.~Tramer, V.~Atlidakis, R.~Geambasu, D.~Hsu, J.-P. Hubaux, M.~Humbert,
  A.~Juels, and H.~Lin, ``Fairtest: Discovering unwarranted associations in
  data-driven applications,'' \emph{EuroS\&P17}, Apr.

\bibitem{breiman2017classification}
L.~Breiman, \emph{Classification and regression trees}.\hskip 1em plus 0.5em
  minus 0.4em\relax Routledge, 2017.

\bibitem{xia2018hyperparameter}
T.~Xia, R.~Krishna, J.~Chen, G.~Mathew, X.~Shen, and T.~Menzies,
  ``Hyperparameter optimization for effort estimation,'' \emph{arXiv preprint
  arXiv:1805.00336}, 2018.

\bibitem{osman2017hyperparameter}
H.~Osman, M.~Ghafari, and O.~Nierstrasz, ``Hyperparameter optimization to
  improve bug prediction accuracy,'' in \emph{MaLTeSQuE}.\hskip 1em plus 0.5em
  minus 0.4em\relax IEEE, 2017.

\bibitem{bias_systemetic}
\BIBentryALTinterwordspacing
J.~Martin, ``Bias or systematic error (validity),'' 2010. [Online]. Available:
  \url{https://www.ctspedia.org/do/view/CTSpedia/BiasDefinition}
\BIBentrySTDinterwordspacing

\bibitem{IBM}
R.~Bellamy, K.~Dey, M.~Hind, S.~C.~Hoffman, S.~Houde, K.~Kannan, P.~Lohia,
  J.~Martino, S.~Mehta, A.~Mojsilovic, S.~Nagar, K.~Natesan~Ramamurthy,
  J.~Richards, D.~Saha, P.~Sattigeri, M.~Singh, R.~Kush, and Y.~Zhang, ``Ai
  fairness 360: An extensible toolkit for detecting, understanding, and
  mitigating unwanted algorithmic bias,'' 10 2018.

\bibitem{biedenkapp2018hyperparameter}
A.~Biedenkapp, K.~Eggensperger, T.~Elsken, S.~Falkner, M.~Feurer, M.~Gargiani,
  F.~Hutter, A.~Klein, M.~Lindauer, I.~Loshchilov \emph{et~al.},
  ``Hyperparameter optimization,'' \emph{Artificial Intelligence}, 2018.

\bibitem{franceschi2017forward}
L.~Franceschi, M.~Donini, P.~Frasconi, and M.~Pontil, ``Forward and reverse
  gradient-based hyperparameter optimization,'' \emph{arXiv preprint
  arXiv:1703.01785}, 2017.

\bibitem{bergstra2011algorithms}
J.~S. Bergstra, R.~Bardenet, Y.~Bengio, and B.~K{\'e}gl, ``Algorithms for
  hyper-parameter optimization,'' in \emph{NIPS11}, pp. 2546--2554.

\bibitem{bergstra2012random}
J.~Bergstra and Y.~Bengio, ``Random search for hyper-parameter optimization,''
  \emph{Journal of Machine Learning Research}, 2012.

\bibitem{pelikan1999boa}
M.~Pelikan, D.~E. Goldberg, and E.~Cant{\'u}-Paz, ``Boa: The bayesian
  optimization algorithm,'' in \emph{GECCO99}, 1999, pp. 525--532.

\bibitem{Will_Koehrsen}
W.~Koehrsen, ``A conceptual explanation of bayesian hyperparameter optimization
  for machine learning,'' 2018.

\bibitem{10.1007/978-3-642-25566-3_40}
F.~Hutter, H.~H. Hoos, and K.~Leyton-Brown, ``Sequential model-based
  optimization for general algorithm configuration,'' in \emph{Learning and
  Intelligent Optimization}.\hskip 1em plus 0.5em minus 0.4em\relax Springer
  Berlin Heidelberg, 2011.

\bibitem{8469102}
V.~{Nair}, Z.~{Yu}, T.~{Menzies}, N.~{Siegmund}, and S.~{Apel}, ``Finding
  faster configurations using flash,'' \emph{TSE}, pp. 1--1, 2018.

\bibitem{Kamishima}
T.~Kamishima, S.~Akaho, H.~Asoh, and J.~Sakuma, ``Fairness-aware classifier
  with prejudice remover regularizer,'' in \emph{Machine Learning and Knowledge
  Discovery in Databases}.\hskip 1em plus 0.5em minus 0.4em\relax Springer
  Berlin Heidelberg, 2012.

\bibitem{NIPS2017_6988}
F.~Calmon, D.~Wei, B.~Vinzamuri, K.~Natesan~Ramamurthy, and K.~R. Varshney,
  ``Optimized pre-processing for discrimination prevention,'' in \emph{NIPS17}.

\bibitem{Hardt}
M.~Hardt, E.~Price, and N.~Srebro, ``Equality of opportunity in supervised
  learning,'' 10 2016.

\bibitem{Kamiran2012}
F.~Kamiran and T.~Calders, ``Data preprocessing techniques for classification
  without discrimination,'' \emph{KAIS12}.

\bibitem{Zhang:2018:MUB:3278721.3278779}
B.~H. Zhang, B.~Lemoine, and M.~Mitchell, ``Mitigating unwanted biases with
  adversarial learning,'' ser. AIES '18.\hskip 1em plus 0.5em minus 0.4em\relax
  NY, USA: ACM.

\bibitem{Kamiran:2018:ERO:3165328.3165686}
F.~Kamiran, S.~Mansha, A.~Karim, and X.~Zhang, ``Exploiting reject option in
  classification for social discrimination control,'' \emph{Inf. Sci.}, 2018.

\bibitem{berk2017convex}
R.~Berk, H.~Heidari, S.~Jabbari, M.~Joseph, M.~Kearns, J.~Morgenstern, S.~Neel,
  and A.~Roth, ``A convex framework for fair regression,'' 2017.

\end{thebibliography}

\end{document}